\documentclass[10pt,a4paper]{article}
\usepackage{amsmath,amssymb,amsfonts}
\usepackage{latexsym}
\usepackage[dvips]{graphicx}
\usepackage{bm}

\newcommand{\R}{\mathbb{R}}                  % the reals
\newcommand{\wh}[1]{\widehat{#1}}
\newcommand{\comm}[2]{\left[#1,#2\right]} % commutator
\DeclareMathOperator{\one}{\mathbb{I}}

\begin{document}

\title{The no-boundary measure in scalar-tensor gravity}

\title{\textbf{The no-boundary measure in scalar-tensor gravity}}
\author{\textsc{Dong-il Hwang}$^{a}$\footnote{dongil.j.hwang@gmail.com},\;\; \textsc{Hanno Sahlmann}$^{b,c}$\footnote{sahlmann@apctp.org}\;\;
and \textsc{Dong-han Yeom}$^{a,d,e}$\footnote{innocent.yeom@gmail.com}\\
\textit{$^{a}$\small{Department of Physics, KAIST, Daejeon 305-701, Republic of Korea}}\\
\textit{$^{b}$\small{Asia Pacific Center for Theoretical Physics, Pohang 790-784, Republic of Korea}}\\
\textit{$^{c}$\small{Physics Department, Pohang University for Science and Technology, Pohang 790-784, Republic of Korea}}\\
\textit{$^{d}$\small{Center for Quantum Spacetime, Sogang University, Seoul 121-742, Republic of Korea}}\\
\textit{$^{e}$\small{Research Institute for Basic Science, Sogang University, Seoul 121-742, Republic of Korea}}}
\maketitle

\begin{abstract}
In this article, we study the no-boundary wave function in scalar-tensor gravity with various potentials for the non-minimally coupled scalar field. Our goal is to calculate probabilities for the scalar field -- and hence the effective gravitational coupling and cosmological constant -- to take specific values. Most calculations are done in the minisuperspace approximation, and we use a saddle point approximation for the Euclidean action, which is then evaluated numerically.

We find that for potentials that have several minima, none of them is substantially preferred by the quantum mechanical probabilities. We argue that the same is true for the stable and the runaway solution in the case of a dilaton-type potential. Technically, this is due to the inclusion of quantum mechanical effects (fuzzy instantons).

These results are in contrast to the often held view that vanishing gravitation or cosmological constants would be exponentially preferred in quantum cosmology, and they may be relevant to the cosmological constant problem and the dilaton stabilization problem.
\end{abstract}
{\vspace{28pt plus 10pt minus 18pt}
     \noindent{\small\rm Preprint no.\ APCTP-Pre2011-004\par}}

\maketitle

\newpage

\section{Introduction}
One of the big challenges of contemporary physics is to obtain a working theory of quantum gravity. But even with such a theory in hand there could presumably remain questions of why the initial conditions of the  gravitational field where what they were. Such questions may even pertain to constants of nature and the like, in a theory of everything, including gravity. It is therefore interesting to look for ways in which the initial conditions can be specified in a \emph{natural} way.

With this in mind, in the present article we study quantum Brans-Dicke theory \cite{brdi01}. In this theory, gravitational coupling and cosmological constant can be dynamical: With a non-trivial potential, one can typically find solutions of the classical equations in which the scalar $\Phi$ carries out small oscillations around a minimum in the potential. Such a minimum can then be thought of as determining a pair of effective constants $(G,\Lambda)$.
We note that a Brans-Dicke theory with potential can be obtained from string theory upon introduction of the dilaton field \cite{Fujii:2003pa,Faraoni:2004pi}. The dilaton field is then related to coupling parameters of all interactions in string theory, and questions about the value of the gravitational coupling strength are thus connected to the \textit{dilaton stabilization problem}, i.e., the problem to explain why the field value of the dilaton of our universe is located in such a stable location \cite{Brustein:1992nk}. This is a problem that has currently no commonly accepted answer.

Upon quantization of the theory, each quantum state gives probabilities for the values of $\Phi$, and these can, provided the state allows for a suitable classical interpretation, be translated into probabilities for effective constants $G$ and $\Lambda$. Such quantum states are obviously not unique, but there are proposals to single out a unique natural state. Given such a proposal, one obtains an a priori probability distribution for  $G$ and $\Lambda$.

For the canonical formulation of quantum gravity, the no-boundary (or Hartle-Hawking) wave function $\Psi_{\text{HH}}$ \cite{Kiefer,Hartle:1983ai,Gibbons:1994cg} is such a proposal for a natural initial state of the universe.\footnote{It should be noted that there are obvious questions regarding how to interpret amplitudes and probabilities calculated from a wave function of the universe such as $\Psi_{\text{HH}}$. There is an extensive literature about and range of approaches to this question, but we will not address any of this here. Rather, we will take a pragmatic approach, by just assuming that the extreme cases (probabilities very large or small, similar or very different) do tell us something about the kind of space times that are excluded, or, respectively, described by a specific model.}

Probabilities for the values of coupling constants have been discussed in the context of quantum gravity for a long time. Early works (see for example \cite{Hawking:1984hk}) rely on the fact that in the instanton approximation, the no-boundary wave-function is dominated by the Euclidean de-Sitter solution, whose action comes out proportional to $1/\Lambda$, and argue that a vanishing cosmological constant is preferred by the resulting probabilities. A famous and detailed argument by Coleman \cite{Coleman:1988tj} has
\begin{eqnarray}
S_{\mathrm{eff}} \propto - \frac{1}{\Lambda G^{2}}
\end{eqnarray}
where $\Lambda$ is to be interpreted as an effective constant that comprises contributions from a path integral over wormhole solutions, and the exponential of (minus) the effective action is to be interpreted as a probability distribution for $\Lambda$. This result has also be interpreted as driving the gravitational coupling to zero \cite{Preskill:1988na}.

Coleman's argument has been studied in the context of scalar-tensor theory \cite{Garay:1992ej}, with the same result for $\Lambda$, and the additional one that the whole theory would be driven toward the Einstein gravity limit.

In the present work, we are more moderate and more ambitious at the same time. We do not consider the sum over topologies, and thus the coupling constants in our action will not receive corrections. On the other hand, as described above, the scalar determines the gravitational coupling, and introducing a potential, we obtain dynamical vacuum energy. Using the no-boundary wave function, we obtain probability distributions for both. We work in the minisuperspace setting \cite{Kiefer,Hartle:1983ai,Gibbons:1994cg,Vilenkin:1986cy}. This has been considered previously for Brans-Dicke theory, for example in  \cite{Liu:1988ri, Zhu:1992zx,Kiefer:1993gj,Lidsey:1995vg,Lidsey:1995ft,Zhu:1998jt,Lidsey:2000rb,Zhu:2000jv}. In \cite{Zhu:1998jt,Zhu:2000jv} various initial conditions for the wave function are studied in regards to the consequences for the effective cosmological constant, using a WKB approximation. The standard result of a minimal value for the cosmological constant is obtained for the tunneling proposal, but interestingly, not necessarily in the no-boundary proposal. What is new in our work is that we strive for a much better approximation to the no-boundary wave function. We follow \cite{Hartle:2008ng,Hartle:2010vi} in allowing fuzzy (i.e., complex-valued) instantons to contribute to the approximation, and we also use numerics to determine the solutions in the fully dynamical regime. The classicality condition of \cite{Hartle:2008ng,Hartle:2010vi} also plays a pivotal role.

In Section~\ref{sec:pre}, we introduce the general procedure to obtain an approximation to the no-boundary measure: Minisuperspace approximation, saddle point approximation of the no-boundary wave function, and imposition of classicality conditions. In Section~\ref{sec:qua}, we apply the method to scalar-tensor gravity, and describe the numerical algorithm that we use to find the relevant stationary points of the Euclidean action.
In Section~\ref{sec:nbm}, we report the results on the no-boundary measure for two types of potentials: quadratic potentials and double or multiple-well potential. Finally, we will apply our results to the dilaton stabilization problem. We have also included analytic results wherever we could. In Section~\ref{sec:dis}, we summarize and discuss our findings.

\section{\label{sec:pre}The no-boundary proposal in quantum cosmology}

\subsection{Canonical quantum cosmology}
\label{se:qcos}
By reducing the full action to homogeneous and isotropic space-times, and upon choosing an appropriate time coordinate $t$
the line element can be written
\begin{eqnarray}
ds^2=N^2(t)dt^2-h_{ij}dx^i dx^j
%\label{eq:}
\end{eqnarray}
where the positive definite metric $h$ only depends on finitely many parameters $q^I(t)$, due to the symmetry assumptions. The corresponding action can typically be brought into the minisuperspace form
\begin{eqnarray}
S[N,q]
=k\int d\lambda\,
{N}\left[\frac{1}{2}G_{AB}
\left(\frac{1}{{N}}\frac{d{q}^A}{d\lambda}\right)
\left(\frac{1}{{N}}\frac{d{q}^B}{d\lambda}\right)
-\widetilde{V}(\wh{q})\right]
\label{eq:particle}
\end{eqnarray}
where $q$ and $N$ comprise the configuration variables, $k$ is the coupling constant, $q^{I}, I=1,2,\ldots n$ comprise the metric degrees of freedom, as well as the degrees of freedom of the matter fields, and
$\widetilde{V}$ is a model-dependent effective potential.
$G$ is the DeWitt metric on minisuperspace.

Let us describe the canonical formulation based on Equation~(\ref{eq:particle}). The canonical momenta conjugate to $q$are given by $p_A=G_{AB}\dot q^B/N$.
$N$ is non-dynamical. Its variation imposes an additional constraint
\begin{eqnarray}
C\equiv\frac{1}{2}G^{AB}p_Ap_B+\widetilde{V}(q)=0
\label{eq:constraint}
\end{eqnarray}
on the variables $\{p_A,q^A\}$. The Hamiltonian $H=NC$ vanishes identically on the constraint hypersurface. %As with all reparametrization invariant systems, time evolution is gauge evolution.
While the kinematical phase space spanned by $\{p_A,q^A\}$ is $2n$-dimensional, after going to the constraint hypersurface $C=0$ and identifying the gauge orbits generated by $C$, we end up with a $2n-2$ dimensional physical phase space.
The system can be quantized \textit{a la} Dirac, with commutation relations
\begin{eqnarray}
\comm{q^A}{p_B}=i\hbar k \,\delta^A_B \, \one.
\end{eqnarray}
Upon choosing a suitable ordering, the constraint $C$ can be turned into an operator, and physical states are required to satisfy the Wheeler-DeWitt equation $C\Psi=0$.
Solutions to this equation are typically not square integrable (zero is in the continuous spectrum of the constraint operator), but if $C$ is selfadjoint, it determines a scalar product on physical states. In practice, it may be very difficult to obtain this scalar product explicitly.

\subsection{Classicality condition}

Let us assume for the discussion that follows, that the physical Hilbert space is a proper subspace of $\mathcal{H}$, and consider a physical state $\Psi(q)$.
Of particular interest are regions in configuration space over which $\Psi$ has a semi-classical form
\begin{eqnarray}
\Psi(q)\approx A(q)e^{i S(q)}, \qquad S(q) \in \R,
\label{eq:class}
\end{eqnarray}
where the rate of change of $S$ is much greater than that of $A$,
\begin{eqnarray} \label{eqn:classicality}
|\nabla_I A(q)|\ll |\nabla_I S(q)|, \qquad I=1,\ldots n.
\end{eqnarray}
One way to see why a state of this form really describes almost classical behavior is the following:
The Wigner function $W[\Psi]$ of $\Psi$ of a state satisfying Equations~(\ref{eq:class}) and (\ref{eqn:classicality}) is approximately
\begin{eqnarray}
W[\Psi](q,p)
\sim |A(q)|^2\,\delta(p-\nabla S).
\end{eqnarray}
This shows that for $q$ in this region, $\Psi$ determines a probability, and a momentum value $p=\nabla S$ which has a high likelihood. Let us consider a surface $\mathcal{S}$ in minisuperspace that is intersected by each trajectory exactly once. Then the probability distribution on this surface in terms of $q$ turns out to be (cf.\ ex.\ \cite{Hartle:2008ng})
\begin{eqnarray}
\rho(q)\approx |A(q)|^2 \,\,n\cdot \nabla S \equiv n \cdot J
\end{eqnarray}
where $n$ is the normal to $\mathcal{S}$ in minisuperspace and $J$ is the conserved ``Klein-Gordon'' current
\begin{eqnarray}
J=-\frac{i\hbar}{2}\Psi^*\overset{\leftrightarrow}\nabla\Psi.
\end{eqnarray}

\subsection{No-Boundary wave function and steepest descent approximation}

\label{se:steepest}
In the minisuperspace approximation, the no-boundary wave function takes the form
\begin{eqnarray}
\Psi_{\text{HH}}(q)=\int_{\text{NB}(q)} \mathcal{D}q(\cdot) e^{-S_{\mathrm{E}}[q(\cdot)]/\hbar},
\label{eq:minhh}
\end{eqnarray}
where $\text{NB}(q)$ are symmetric Euclidean space-time histories $q(\cdot)$ with a single boundary, the geometry of which is described by $q$.
We assume that the parametrization of the histories, or equivalently, the lapse function, has been fixed appropriately. But even after gauge-fixing, the integral is not convergent because the Euclidean action is in general not positive. One way to possibly cure this divergence is to regard the path integral in Equation~(\ref{eq:minhh}) as a contour integral, and deform the path away from real superspace, into the space of \emph{complex} symmetric metrics. Despite the huge simplification due to the minisuperspace approximation, the path integral (Equation~(\ref{eq:minhh})) is hard to treat exactly. Therefore one uses the steepest descent approximation.
One obtains\footnote{Note that, here and in the following, $S_{\mathrm{E}}[q]$ will sometimes be considered a \emph{function} of the \emph{endpoints} of the history $q(\cdot)$. We hope that this does not lead to confusion.}
\begin{eqnarray}
\Psi_{\text{HH}}(q)
\approx \sum_{\text{ext}}
P(q_\text{ext})e^{-S_{\mathrm{E}}[q_\text{ext}]/\hbar}
\end{eqnarray}
where $q_\text{ext}$ is an extremizing Euclidean history satisfying the appropriate (no-)boundary conditions, and $P$ is given by
1 plus higher order terms which depend on functional derivatives of the action. We will neglect the higher orders of $P$ in what follows. For a single extremum (in the application below we will see that the boundary conditions allow at most two extrema) we then have\footnote{In this paper, $\mathfrak{Re}$ denotes the real part and $\mathfrak{Im}$ denotes the imaginary part.}
\begin{eqnarray}
\Psi_{\text{HH}}(q)
\approx
e^{-S_\text{E}[q_\text{ext}]/\hbar}=e^{-\mathfrak{Re} S_{\text{E}}/\hbar}e^{-i\mathfrak{Im} S_{\text{E}}/\hbar}.
\label{eq:hhapprox}
\end{eqnarray}
Comparing with Equation~(\ref{eq:class}), we see that $\Psi_{\text{HH}}$ is of semiclassical form at $q$ if
\begin{eqnarray}
|\nabla_I \mathfrak{Re} S_{\,\mathrm{E}}| \ll |\nabla_I \mathfrak{Im} S_{\,\mathrm{E}}|.
\end{eqnarray}
Let us finish by discussing a subtle point that is nevertheless important in applications: we have said earlier that to give meaning to the path integral, we have to gauge-fix the lapse, or equivalently, the time parameter used in the action. A priori, the time parameter is real, but it is very convenient to also consider \emph{complex} time parameters. In fact, if the Lagrange function and the history are analytic in suitable regions, the action can be viewed as a contour integral in the complex time plane, and hence depends on the integration contour only through its endpoints. Following \cite{Hartle:2007gi,Hartle:2008ng} we will make use of the freedom to chose the contour, below. Since we have written the no-boundary wave function in terms of the Euclidean action, we will refer to a section of a contour that is parallel to the real axis as \emph{Euclidean}, to one that is parallel to the imaginary axis as \emph{Lorentzian}.

\section{\label{sec:qua}The no-boundary measure in scalar-tensor gravity: Formalism}

In this section, we investigate the no-boundary measure for scalar-tensor gravity. Scalar-tensor gravity is of interest not only in itself, but also as a dilaton gravity limit of string theory.
For us, it is interesting because it has dynamic gravitational coupling and vacuum energy.
Quantization of scalar-tensor gravity has certainly been considered before. In the context of quantum cosmology see for example  \cite{Liu:1988ri, Zhu:1992zx,Kiefer:1993gj,Lidsey:1995vg,Lidsey:1995ft,Zhu:1998jt,Lidsey:2000rb,Zhu:2000jv} and our discussion in the introduction. For non-perturbative quantization, see for example \cite{Zhang:2011vg,Zhang:2011qq,Zhang:2011vi,Qiang:2009fu}.

The first and simplest scalar tensor theory is that of Brans and Dicke \cite{brdi01},
\begin{eqnarray}
\label{eq:bd}
S = \frac{1}{16\pi} \int \sqrt{-g} d^4 x \left( \Phi R - \omega g^{\alpha\beta} \frac{\nabla_{\alpha}\Phi
\nabla_{\beta}\Phi}{\Phi} \right).
\end{eqnarray}
Here, $\omega$ is a dimensionless coupling parameter, and Einstein gravity is restored in the $\omega \rightarrow \infty$ limit. We will refer to more general actions involving a non-minimally coupled scalar as \textit{scalar-tensor theories}, whereas we reserve the name \textit{Brans-Dicke theory} to the theory with the above action, possibly with the inclusion of a potential $V(\Phi)$.
Matching current observations against the Brans-Dicke theory without a potential requires $\omega$ greater than $\sim 40,000$ \cite{Ber}. Small values for $\omega$ are however admissible for non-trivial potentials
\cite{Khoury:2003aq}, and can be found in string-inspired models \cite{Fujii:2003pa,Faraoni:2004pi}.

We briefly comment on possible scenarios that involve Brans-Dicke theory.
The most important one is dilaton gravity obtained from string theory. The effective action of string theory has the form \cite{Gasperini:2007zz}:
\begin{eqnarray}
\label{eq:dilaton} S = \frac{1}{2 \lambda_{s}^{d-1}}\int d^{d+1}x \sqrt{-g} e^{-\phi} \left( R + (\nabla \phi)^{2} \right),
\end{eqnarray}
with $d$ the space dimension, $\lambda_{s}$ the length scale of string units, $R$ the Ricci scalar, and $\phi$  the dilaton field. The field redefinition $\Phi/(8 \pi G_{d+1})\equiv\exp(-\phi)/\lambda_{s}^{d-1}$ turns the above action into that of Brans-Dicke theory with $\omega = -1$.

In the first model of Randall and Sundrum \cite{Randall:1999ee}, two branes have been employed to account for the hierarchy problem. Due to the warp factor between two branes, one obtains a positive tension brane and a negative tension brane in the anti de Sitter space background. According to Garriga and Tanaka \cite{Garriga:1999yh} each brane can be described by Brans-Dicke theory in the weak field limit with
\begin{eqnarray}
\label{eq:gt} \omega = \frac{3}{2} \left( e^{\pm s/l} - 1 \right),
\end{eqnarray}
where $s$ is the location of the negative tension brane along the fifth dimension, $l=\sqrt{-6/\Lambda}$ is the length scale of the anti de Sitter space, and the sign $\pm$ denotes the sign of the tension. To explain the hierarchy problem, we require $s/l \sim 35$. We then obtain a sufficiently large value of $\omega$ on the positive tension brane while $\omega \gtrsim -3/2$ on the negative tension brane \cite{Garriga:1999yh,Fujii:2003pa}. In principle, however, $s/l$ can be chosen arbitrarily, and hence one may infer that various $\omega$ near $-3/2$ may be allowed by models of the brane world scenarios.

Finally, it is well known that $f(R)$-gravity (for a recent review see \cite{Sotiriou:2008rp}) can be reformulated in terms of a Brans Dicke field.  The action is
\begin{eqnarray}
S = \int dx^{4} \sqrt{-g} \left[ \frac{1}{16\pi} f(R) + \mathcal{L}_{\mathrm{matter}} \right].
\end{eqnarray}
Introducing an auxiliary field $\psi$, we change the gravity sector to
\begin{eqnarray}
S_{\mathrm{gravity}} = \frac{1}{16\pi} \int dx^{4} \sqrt{-g} \left[f(\psi) + f'(\psi) (R-\psi) \right]
\end{eqnarray}
with constraint $\psi = R$. If we define a new field $\Phi$ by $\Phi = f'(\psi)$, we obtain
\begin{eqnarray}
S_{\mathrm{gravity}} = \frac{1}{16\pi} \int dx^{4} \sqrt{-g} \left[\Phi R - V(\Phi) \right]
\end{eqnarray}
where $V(\Phi) = - f(\psi) + \psi f'(\psi)$.
This is exactly the $\omega = 0$ limit of the Brans-Dicke theory.

For our purposes, we begin from the action (Equation~(\ref{eq:bd})) by adding a potential $V(\Phi)$.
In this article, we focus on the two types of potentials: quadratic potentials
\begin{eqnarray}
V(\Phi) = \frac{1}{2} M^{2} (\Phi - 1)^{2} = \frac{1}{2} M^{2} \Xi^{2},
\end{eqnarray}
where $\Xi=\Phi-1$, and double-well potentials,
\begin{eqnarray}
V(\Phi) = \Phi^{2} \left( \int_{1}^{\Phi} \frac{F(\bar{\Phi})}{\bar{\Phi}^{3}} d \bar{\Phi} + V_{0} \right)
\end{eqnarray}
where we use an effective force function $F(\Phi)$ given by\footnote{Note that the potential in the Einstein frame $\hat{V}$ is \cite{Faraoni:2004pi}
\begin{eqnarray}
\hat{V}(\Phi) = \int_{1}^{\Phi} \frac{F(\bar{\Phi})}{\bar{\Phi}^{3}} d \bar{\Phi} + V_{0}.
\end{eqnarray}
The dynamical field $\hat{\Phi}$ in that frame is related to $\Phi$ by $\Phi = \exp \hat{\Phi}[(16 \pi)/(2\omega+3)]^{1/2}$.  Therefore, the only effect is to stretch the potential along the field direction, and this does not affect the vacuum energy of each field value.}
\begin{eqnarray}
F(\Phi) \equiv \Phi V'(\Phi) - 2 V(\Phi)
= A \left(\Phi-\Phi_{\mathrm{a}} \right) \left(\Phi-\Phi_{\mathrm{b}} \right) \left(\Phi- \left( \frac{\Phi_{\mathrm{a}}+\Phi_{\mathrm{b}}}{2} + \delta \right) \right).
\end{eqnarray}
$A$ is a positive constant, $\Phi_{\mathrm{a}}$ and $\Phi_{\mathrm{b}}$ denote the field values of each vacuum, and $\delta$ is a free parameter that determines the location of the bump of the potential.
For convenience, we choose $\Phi_{\mathrm{a}}=1$ and let $V(\Phi_{\mathrm{a}})=V_{0}$ be true vacuum value. It is also useful to introduce the effective potential
\begin{eqnarray}
U(\Phi) = \int_{1}^{\Phi} F(\bar{\Phi}) d \bar{\Phi} = \int_{1}^{\Phi}\left( \bar{\Phi} V'(\bar{\Phi}) - 2V(\bar{\Phi}) \right) d \bar{\Phi},
\end{eqnarray}
which is the one showing up in the field equation, $\nabla^{2} \Phi = U'/(3+2\omega)$.

In this article, we will not explicitly perform calculations for a dilaton type potential resulting from the field redefinition in Equation~(\ref{eq:dilaton}), but conduct a qualitative discussion of the stability issue based on an approximation by a double well potential. For a quantitative study of the dilaton type potentials see \cite{next}.

We write the symmetric line elements as\footnote{For some more detail regarding the following calculations see for example \cite{Kim:2010yr}.
}
\begin{eqnarray}
ds_{\mathrm{L,E}}^{2} = \mp N^{2}(t) dt^{2} + \rho^{2}(t) (d\chi^{2} + \sin^{2} \chi (d\theta^{2} + \sin^{2}\theta d\varphi^{2})),
\end{eqnarray}
and obtain the symmetry reduced actions
\begin{eqnarray}
\begin{split}
S_\text{L,E}[\rho, \Phi, N]
=\frac{\pi}{8} \int dt\, N&\left(
\pm 6\rho^{2} \dot\Phi\dot\rho N^{-2}
\pm 6\Phi\rho\dot{\rho}^2N^{-2}\right.\\
&\left. \qquad -6\Phi\rho
\mp \omega\rho^3\Phi^{-1}\dot{\Phi}^2N^{-2}
+\rho^3 V(\Phi)
\right)
\end{split}
\label{eq:lac}
\end{eqnarray}
To obtain the above result, we have added a suitable total derivative to keep the action differentiable in the presence of boundaries. $dt$ should be changed to $d\eta$ for Euclidean case.

For the approximation of the no-boundary wave function, the on-shell Euclidean action and both, Euclidean and Lorentzian equations of motion are needed. Using the Euclidean Einstein equation for $\rho$ and field equation for $\phi$ (with $N=1$),
one finds
\begin{eqnarray}
S_{\,\mathrm{E}} = \frac{\pi}{4} \int d \eta \left( \rho^{3} V - 6 \rho \Phi \right).
\end{eqnarray}

The equations of motion (with $N=1$) for the Euclidean time $\eta$ and the Lorentzian time $t$ are as follows:
\begin{eqnarray}
\label{BD1}\ddot{\Phi} &=& - 3 \frac{\dot{\rho}}{\rho} \dot{\Phi} \pm \frac{1}{2 \omega + 3} \left(\Phi V' - 2V\right), \\
\label{BD2}\ddot{\rho} &=& \frac{\pm 1 - \dot{\rho}^{2}}{\rho} - \omega \rho \frac{\dot{\Phi}^{2}}{6 \Phi^{2}} \mp \frac{\rho V'}{4 \omega + 6} \mp \frac{2 \omega \rho}{6 \omega + 9} \frac{V}{\Phi},
\end{eqnarray}
where the upper signs are for Euclidean and the lower signs are for Lorentzian.
Note that all functions in the equations are complex in general ($\rho=\rho^{\mathfrak{Re}}+i\rho^{\mathfrak{Im}}$ and $\Phi=\Phi^{\mathfrak{Re}}+i\Phi^{\mathfrak{Im}}$). Therefore, effectively, there are four functions ($\rho^{\mathfrak{Re}}$, $\rho^{\mathfrak{Im}}$, $\Phi^{\mathfrak{Re}}$, $\Phi^{\mathfrak{Im}}$) and we need eight initial conditions to fix a solution.

As we have already sketched the general procedure in Section~\ref{se:qcos}, we can be brief here. The symmetry reduced Lorentzian action, Equation~(\ref{eq:lac}), is of the form Equation~(\ref{eq:particle}),
where $q=(\rho, \Phi)$ and $N$ comprise the configuration variables, the coupling constant takes the value $k=\pi/8$, the effective potential is given by $\widetilde{V}=6\rho\Phi-\rho^3 V$, and
\begin{eqnarray}
G_{\rho\rho}=12\rho\Phi, \qquad G_{\Phi\rho}=G_{\rho\Phi}=6\rho^2, \qquad G_{\Phi\Phi}=-2\omega\frac{\rho^3}{\Phi}
%\label{eq:}
\end{eqnarray}
are the components of the DeWitt metric on minisuperspace. Thus the system is equivalent to a relativistic particle propagating in a two-dimensional space with metric $G_{AB}$ under the influence of a potential $\widetilde{V}$. The determinant of the metric is $-12(2\omega+3)\rho^4$, so $G_{AB}$ is Lorentzian for $\omega>-3/2$ and Euclidean for  $\omega<-3/2$.

The detailed expressions for the momenta and the Hamilton constraint $C$ do not concern us here, but let us make some remarks on the reduced phase space.
As with all reparametrization invariant systems, time evolution in gauge evolution. While the kinematical phase space spanned by $\{p_A,q^A\}$ is $4$-dimensional, after going to the constraint hypersurface $C=0$ and identifying the gauge orbits generated by $C$, we end up with a two-dimensional physical phase space. Points in this space are simply the trajectories $\Phi(\rho)$ that can be obtained by solving Equation~(\ref{eq:constraint}) to obtain $q^A(\lambda)$ and then eliminate $\lambda$ which is possible at least locally.

Let us also give a parametrization of the space of trajectories $\Phi(\rho)$. Let us assume that the particle is not a tachyon -- the parametrization can also be easily adapted to the more general case, but details then depend on the potential. We fix an initial value slice $\rho=\rho_0=\text{const}$. Note that this slice is spatial with respect to $G$. Then each solution of the dynamics will intersect this slice once, and we obtain $\Phi(\rho_0)$ and
the two momenta $p_A(\rho_0)$ at the intersection point. We note that reparametrization of the solution changes the momenta by a constant factor, so we can chose as parameters $\Phi(\rho_0)$ and the ratio $p_1(\rho_0)/p_2(\rho_0)$.

\subsection{Steepest descent approximation}
We remind the reader that to obtain the steepest descent approximation of the no-boundary wave function (see Section~\ref{se:steepest}) at a superspace point $\rho,\Phi$, we need to determine complex solutions of the equations of motion that have $\rho$ and $\Phi$ as boundary values and otherwise fulfill the no-boundary condition. To specify the boundary value problem fully, and to compute the action of such a solution, we need to fix a contour in the complex time plane. While this choice is in principle largely arbitrary, there are practical reasons to chose one contour over the other. To formulate the no-boundary condition, it is necessary to start the contour as Euclidean. Moreover, the interpretation of the results is done most easily if the end of the contour is Lorentzian.

The contour we chose for the practical evaluation consists of two components:
An Euclidean component from $\eta = 0$ to $\eta = X$ and a Lorentzian component from $\eta = X$ to $\eta = X + i Y$ (or from $t=0$ to $t=Y$ with $\eta = X + it$). To implement the no-boundary condition, we require that
\begin{eqnarray}
\rho(0)^{\mathfrak{Re}} = \rho(0)^{\mathfrak{Im}} = 0,&\qquad&
\dot{\rho}(0)^{\mathfrak{Re}} = 1,\quad
\dot{\rho}(0)^{\mathfrak{Im}} = 0,\\
\dot{\Phi}(0)^{\mathfrak{Re}} = \dot{\Phi}(0)^{\mathfrak{Im}} = 0.&&
\end{eqnarray}

At the turning point $\eta = X$, we have to match the Euclidean part of solutions (so called, $\rho$, $\dot{\rho}$, $\Phi$, $\dot{\Phi}$) to the Lorentzian part of solutions (so called, $\underline{\rho}$, $\underline{\dot{\rho}}$, $\underline{\Phi}$, $\underline{\dot{\Phi}}$) of Equations~(\ref{BD1}) and (\ref{BD2}) via the matching conditions:
\begin{eqnarray}
\underline{\rho}(t=0) = \rho(\eta=X), &\qquad& \underline{\dot{\rho}}(t=0)=i\dot{\rho}(\eta=X),\\
\underline{\Phi}(t=0) = \Phi(\eta=X), &\qquad& \underline{\dot{\Phi}}(t=0)=i\dot{\Phi}(\eta=X),
\end{eqnarray}
since $d/dt = i d/d\eta$.

With a view to the calculation of the no-boundary wave function, $\Psi(\rho,\Phi)$ we note that we have $6$ real conditions at $\eta=0$, which leaves two real initial values to be specified. Additionally, our contour has two parameters $(X,Y)$. This makes $4$ real parameters that we can shoot for at the boundary, i.e., precisely the number needed to tune the arguments of the
no-boundary wave function. We are, however, only interested in regions of configuration space on which the wave function has semiclassical form. This is what we discuss next.

\subsection{Classicality condition}
We are not interested in the value of the no-boundary wave function for arbitrary arguments, but only in regions of configuration space on which the wave function has semiclassical form Equations~(\ref{eq:class}) and (\ref{eqn:classicality}). We note that the latter condition actually has two components, of which one can be checked easily in our setup. The easy component is the one in the direction of time evolution:
Let $q(t)\equiv (a(t),\Phi(t))$ one of the no-boundary histories, evaluated on part of the Lorentzian contour. With $\delta q=dq/dt\rvert_{t_0}$, we find
\begin{eqnarray}
 \delta q^A \nabla_A S_\text{E}(q(t_0))= \left. \frac{d}{dt}\right\rvert_{t_0} \int^t \mathcal{L}(q(t'),\dot q(t'))\,dt'  =
\mathcal{L}(q(t_0),\dot q(t_0)),
\label{eq:nas}
\end{eqnarray}
where $\mathcal{L}$ is the Lagrange function.
Thus, given a history $q(t)$, the component of $\nabla_A S_\text{E}$ \emph{in the direction of} $\delta q$ is easy to determine. We also note that, by definition, in a classical region $q^A$ and $\delta q^A$ have to be real.
In what follows, and in slight abuse of terminology, we will call a no-boundary solution of the equations of motion which satisfies the tangential component of classicality, with $\delta q$ real, a \emph{classical history}.

The orthogonal component of the classicality condition is harder to check, as it involves finding no-boundary solutions of the equations of motion with nearby boundary conditions.

For the numerical evaluation, we split the initial value of $\Phi$ as
\begin{eqnarray}
\Phi(0) = \Phi_{0} e^{i \theta}
\end{eqnarray}
with $\Phi_{0}$ and $\theta$ real. We fix the latter, and then numerically scan the $(\theta,X)$ space for solutions that
\begin{enumerate}
	\item become classical at late times, and
	\item have $\underline{\rho}^{\mathfrak{Im}}$, $\underline{\dot{\rho}}^{\mathfrak{Im}}$, $\underline{\Phi}^{\mathfrak{Im}}$, and $\underline{\dot{\Phi}}^{\mathfrak{Im}}$ approach zero in the large $t$ limit.
\end{enumerate}
The parameter $Y$ is held fixed (and sufficiently large), as it simply corresponds to time evolution in Lorentzian time.

\subsection{Probabilities}
We will denote the space of histories satisfying the no-boundary condition by $\mathfrak{H}$.\footnote{Note that histories which differ by time reparametrization are considered the same. Notice also that while these histories are regular in Euclidean time as per the no-boundary conditions, they may well have singularities along the real time axis.} We eventually want to calculate the probability of histories satisfying certain (boundary-) conditions. Let us say that we are interested in histories satisfying a certain condition $A$. Then we can define the subset  by
\begin{eqnarray}
\mathfrak{H}_{A} = \{h \in \mathfrak{H} \;|\; h \mathrm{\, has\, property\,} A\}
\end{eqnarray}
of $\mathfrak{H}$.
Given a set of classical histories $\mathfrak{H}_A$, in principle we have
\begin{eqnarray}
P_A=\int_{Q_A} |\Psi_{\text{HH}}(h,\chi)|^2 n \cdot \nabla S \, \mathcal{D}\mu(h,\chi).
%\label{eq:}
\end{eqnarray}
The integration is over a subset $Q_A$ of a spatial slice with normal $n$ in superspace.
$Q_A$ is the set of points on this slice such that the wave function satisfies the classicality condition in a way compatible with the condition $A$. $S$ was defined in Equation~(\ref{eq:class}). $\mu$ is a certain measure which can be obtained in principle from the inner product on the space of solutions to the Wheeler-DeWitt equation and the slice. But it is very difficult to obtain in practice. Using minisuperspace and steepest descent approximation, using $\Phi_0$ as parameter on the slice, and ignoring details of the measure as well as the variation of  $n \cdot \nabla S$, by assuming both are constant over the space of histories
\begin{eqnarray}
P_A\approx \frac{1}{Z}\int_{Q_A} |\exp(-S_\text{E}[h_{\Phi_0}])|^2 d\Phi_0=
\frac{1}{Z}\int_{Q_A}e^{-2\mathfrak{Re} S_\text{E}[h_{\Phi_0}]} d\Phi_0
\label{eq:prob}
\end{eqnarray}
where $Z$ is some normalization constant and $h_{\Phi_0}$ is a history that initially has scalar field modulus equal to $\Phi_0$.

If we have two conditions $A$, $B$, where $A$ is an initial and $B$ a final condition, we will use the notation $\mathfrak{H}_{A\rightarrow B}$ for $\mathfrak{H}_A\cap \mathfrak{H}_B$, and $P_{A\rightarrow B}$ for the corresponding probability.

If $S_\text{E}[h_{\Phi_0}]$ is slowly varying, and we compare two probabilities, then from Equation~(\ref{eq:prob}) we get
\begin{eqnarray}
\frac{P_{A_1}}{P_{A_2}}\approx \exp\left(-2S_\text{E}[h_{\overline{\Phi}_1}]+2S_\text{E}[h_{\overline{\Phi}_2}]\right) \frac{\int_{Q_{A_1}} d\Phi_0}{\int_{Q_{A_2}} d\Phi_0},
\end{eqnarray}
where $\overline{\Phi}_1$ is a suitable initial condition such that the history fulfills $A_1$, and $\overline{\Phi}_2$ the same for $A_2$.

\subsection{Searching algorithm}

We have to find initial conditions, the initial phase angle $\theta$, and the turning point $X$, for a given initial field amplitude $\Phi_{0}$ to satisfy the classicality condition. To find classical histories for a given $\Phi_{0}$, we formulate an \textit{optimization problem} and solve it using the idea of the \textit{generic algorithm}. To realize this algorithm, first we define the \textit{objective function} that quantifies classicality. Second, using the optimization algorithm, we list and choose the best candidates of initial conditions using the scores of the objective function. Third, we inversely check whether the solutions of the initial conditions really have classical properties what we required.

We first define the standard objective function $F_{\Phi_{0}}$. To this end, note that the tangential component of classicality can be reformulated as
\begin{eqnarray}
\left\lvert\frac{\delta q^A (\nabla_A S_\text{E})^\mathfrak{Re}}
{\delta q^A \nabla_A S_\text{E}}\right\rvert\equiv \left\lvert\frac{\mathcal{L}^{\mathfrak{Re}}(q,\dot{q})}{\mathcal{L}(q,\dot{q})}\right\rvert\ll 1,
\label{eq:equiclass}
\end{eqnarray}
as long as $\delta q^A$ is real (see the discussion around Equation~(\ref{eq:nas})).
Evaluating the left-hand side of Equation~(\ref{eq:equiclass}) in a single point given by a numerically determined function is not a very stable procedure. Thus, in practice we average it over some time interval, to define the objective function,
\begin{eqnarray}
F_{\Phi_{0}} \left[\theta, X \right] \equiv \int_{T_{1}}^{T_{2}} \left| \frac{\mathcal{L}^{\mathfrak{Re}}_{\Phi_{0}}[\theta, X](t)}{ \mathcal{L}_{\Phi_{0}}[\theta, X](t)}\right| dt.
\end{eqnarray}
$T_{1}$ and $T_{2}$ define the Lorentzian time interval where the classicality will be tested, and need to be chosen sufficiently large. The reality of $\delta q^A$ is checked separately for minima of $F$. Altogether we obtain a criterion that is, for asymptotically large $T$, strictly equivalent to one component of classicality. Other definitions for objective functions that single out classical histories are undoubtedly possible. As said sketched above, ours was chosen for numerical convenience\footnote{We have run checks with other possible definitions for the objective functions. We found that the results generally agree, although some differences are observed in the performance of the numerical search algorithm, in particular in some more extreme regimes.}.

Then, we can define the relevant optimization problem: for given $\Phi_{0}$ and other constraints, which values of initial parameters $(\theta, X)$ minimizes $F_{\Phi_{0}}$? Of course, we can find the optimal solution by searching all possible values of $(\theta, X)$. However, it takes too much time and computation power, and hence we need a better optimization algorithm.

In this paper, we use a simple version of the generic algorithm.
\begin{description}
\item[Initialization:] We first generate a set of initial condition pairs of a number $N$: $\{(\theta_{i}, X_{i})\}$ ($i=1, ..., N$), where we choose $\theta$ and $X$ randomly. This set of $(\theta_{i}, X_{i})$ constitutes the first \textit{generation} of candidate solutions.
\item[Elite:] Then, we calculate $F_{\Phi_{0}}$ for each $(\theta_{i}, X_{i})$ by solving equations of motions numerically. Among the parameters $(\theta_{i}, X_{i})$, the one that has smallest $F_{\Phi_{0}}$ can be regarded as the most classical solution in the first generation. As a subset of candidates, we select the $n_{e} < N$ number of initial conditions that have smallest $F_{\Phi_{0}}$. We call them elites of the first generation.
\item[Cross-over:] We cross-over the elite parameters to generate the $n_{c}$ $(n_{e} + n_{c} < N)$ number of new parameters. That is, we pick two parameters $(\theta_{i}, X_{i})$ and $(\theta_{j}, X_{j})$ for arbitrary chosen $i$ and $j$ from the elites and define a new parameter $((\theta_{i}+\theta_{j})/2, (X_{i}+ X_{j})/2)$; we repeat this process $n_{c}$ times. They are adopted to examine the parameter space around the elite group more detail.
\item[Mutation:] Finally, we generate the $n_{m}$ $(n_{e} + n_{c} + n_{m} = N)$ number of totally new parameters, so-called mutations, by choosing $\theta$ and $X$ randomly. They ensure us that potential candidates of parameters are not lost systematically.
\item[Evolution:] The set of elites, cross-overs, and mutations constitutes the second generation. Using the second generation of initial conditions, we calculate the objective function $F_{\Phi_{0}}$. Then we can choose new elites, cross-overs among the new elites, new mutations, and define the third generation. We repeat this process to evolve generations until the values $\theta$ and $X$ of the elite group reaches to a steady state.
\end{description}
This generic algorithm finds the optimal parameter as the number of generations becomes sufficiently large. However, it is possible that there is no classical solution for a given $\Phi_{0}$. Therefore, we have to be careful whether the optimized value is really the classical solution or not.

There are two main drawbacks of the generic algorithm in searching classical solutions. One is that it can converge to a local minimum of $F_{\Phi_{0}}$ rather than the global minimum, or there can be two or more classical histories for a given $\Phi_{0}$. To resolve this problem, we divided the searching region of $(\theta, X)$ into several pieces and found the optimal parameters separately. In this way, one can reduce the possibility of being captured by a local minimum. The other drawback is that the result can sensitively depend on simulation parameters: $N$, $n_{e}$, $n_{c}$, $n_{m}$, and the definition of the objective function. $N$, $n_{e}$, $n_{c}$ and $n_{m}$ should be sufficiently large until the result is not sensitively depend on the choice of them. In our choice of $F_{\Phi_{0}}$, we have to choose a proper time interval $[T_{1}, T_{2}]$ where we will regard that a history becomes classicalized around the time. Therefore, we have to carefully choose the time interval case by case and check whether the results are not sensitively depend on the choice of parameters.

\section{\label{sec:nbm}The no-boundary measure in scalar-tensor gravity: Results}

In this section, we study the no-boundary measure for two types of potentials: quadratic potentials
and the double-well/multiple-well potentials. The former is useful to obtain clues abut the physics the near a generic local minimum, the latter to compare the probabilities for various values of gravitational couplings
In the case of the quadratic potential, we also have some analytic results that we can compare to the numerical ones. Our results on these types of potentials will give some intuition for the dilaton stabilization problem.
First, we will make some analytic considerations, then we will turn to the numerical investigation.

\subsection{Analytic considerations regarding classical histories}

For a first estimate, we regard that the potential is approximated near $\Phi = 1$ by
\begin{eqnarray}
V(\Phi) \simeq \frac{1}{2} M^{2} (\Phi - 1)^{2} + \mathcal{O}((\Phi-1)^{3}).
\end{eqnarray}
After we re-define the field $\Xi = \Phi -1$, the on-shell Euclidean action becomes
\begin{eqnarray}
S_{\,\mathrm{E}}
= \frac{\pi}{4} \int d \eta \left( \frac{1}{2} \rho^{3} M^{2} \Xi^{2} - 6 \rho \Xi - 6 \rho \right).
\end{eqnarray}
Therefore, compared to the Einstein case, we have an effective new term proportional to $-\rho \Xi$.
The field equation is
\begin{eqnarray}
\ddot{\Xi} = - 3 \frac{\dot{\rho}}{\rho} \dot{\Xi} + m^{2} \Xi,
\end{eqnarray}
where we have defined $m^{2}=M^{2}/(2\omega+3)$. Thus, for a quadratic potential the Brans-Dicke field equation is the same as that for a standard scalar coupled to Einstein gravity. Thus, as it was approximated by \cite{Lyons:1992ua}, we can compare the Euclidean action to the slowly rolling and almost static field limit.

In the static case, where $\Phi=\Phi_0=\text{const.}$ with $\Phi_0$ determined by $\Phi_0 V'(\Phi_0)- 2V(\Phi_0)=0$, we have the exact solution,
\begin{eqnarray}
\label{eq:exact}
\Phi(\eta) = \Phi_{0}, \;\;\; \rho(\eta) = \frac{1}{A} \sin A\eta,
\end{eqnarray}
where
\begin{eqnarray}
A^2=\left[\frac{1}{4\omega+6}+\frac{\omega}{6\omega+9}\right]V' = \frac{1}{6}V'.
\end{eqnarray}
To satisfy no-boundary and classicality conditions, the turning point of the contour should be at $X=A\pi/2$. Then the real part of the action does not grow after the turning point and the action becomes
\begin{eqnarray}\label{eqn:dS}
S_{\,\mathrm{E}} = \frac{\pi}{4} \int^{X}_{0} d \eta \left( \rho^{3} V(\Phi_{0}) - 6 \rho \Phi_{0} \right) = - \frac{3 \pi \Phi_{0}^{2}}{V(\Phi_{0})}.
\end{eqnarray}
Therefore, for small $\Xi$ limit, the Euclidean action negatively increases and hence $\Phi \sim 1$ will be preferred.

Let us go even a little further: We have found one exact solution fulfilling classicality, and it is a de Sitter-like solution, i.e. trigonometric/exponential dependence of the scale factor in Euclidean respectively Lorentzian time. Thus it is not unreasonable to guess that other solutions contributing to the classical regime are perturbed de Sitter like solutions. To obtain those, we need a field configuration allowing an inflating space-time. The well-known mechanisms are slow-roll inflation and false vacuum inflation. We are therefore led to hypothesize that histories contributing to the classical regime are allowed under the following conditions:
\begin{enumerate}
\item A part of the potential satisfies the slow-roll conditions and the history experiences the region of the potential (thus mimicking slow-roll inflation).
\item The field becomes slowed-down and approaches a local maximum or minimum to experience an effectively de Sitter space (thus mimicking false vacuum inflation).
\end{enumerate}
The slow-roll conditions for Einstein gravity are
\begin{eqnarray}
\left( \frac{\hat{V}'}{\hat{V}} \right)^{2} \ll 1, \;\;\;\; \frac{\hat{V}''}{\hat{V}} \ll 1.
\end{eqnarray}
In terms of the Brans-Dicke potential,
\begin{eqnarray}
\label{eqn:SL}
\left(\frac{16 \pi}{2\omega + 3}\right)\left( 2 - \Phi \frac{V'}{V} \right)^{2} \ll 1,
\quad
\left(\frac{16 \pi}{2\omega + 3}\right)\left( 4 - 3 \Phi \frac{V'}{V} + \Phi^{2}\frac{V''}{V} \right) \ll 1.
\end{eqnarray}
From this consideration, one is led to the following hypotheses:
First, assuming the local extrema to be obtained for $|\Phi| \sim \mathcal{O}(1)$, both of the slow-roll conditions are difficult to satisfy, unless there is an inflection point.\footnote{Maybe an exceptional case can happen if the potential is $\sim \Phi^{n} + \Lambda$ and $\Lambda$ is larger than $1$; however, a cosmological constant  larger than $1$, is unrealistic for our universe and we are not interested such a case.}
Second, assuming $V\sim \Phi^{n}$ for large $\Phi$, slow-roll conditions hold only if $n=2$. Therefore, if the potential shape is ``runaway'' for large $\Phi$, i.e., $n < 0$, classicalization via slow-roll inflation may not happen.
Certainly, in both these cases, classical histories may still appear via false vacuum inflation.

For the quadratic potential, the conditions are
\begin{eqnarray}
4 \left(\frac{16 \pi}{2\omega + 3}\right) \left( 1 - \frac{\Phi}{\Phi-1} \right)^{2} \ll 1,\quad
2 \left(\frac{16 \pi}{2\omega + 3}\right) \left( \left(\frac{\Phi}{\Phi-1}\right)^{2} - 3 \frac{\Phi}{\Phi-1} + 2 \right) \ll 1.
\end{eqnarray}
Therefore, both of the conditions hold if and only if in the $\Phi \gg 1$ limit. This suggests that there are no classical histories that start near the bottom of the potential $\Phi_{0} \sim 1$. Such behavior has been observed numerically in the Einstein case \cite{Hartle:2008ng}. It also suggests that the range of initial conditions that allow for classical histories gets larger as $\omega$ increases. Both of these conclusions from our analytical considerations are borne out very well in the numerical results that we will present below.

\subsection{Numerical results on the quadratic potential}
\begin{figure}
\begin{center}
\includegraphics[scale=0.25]{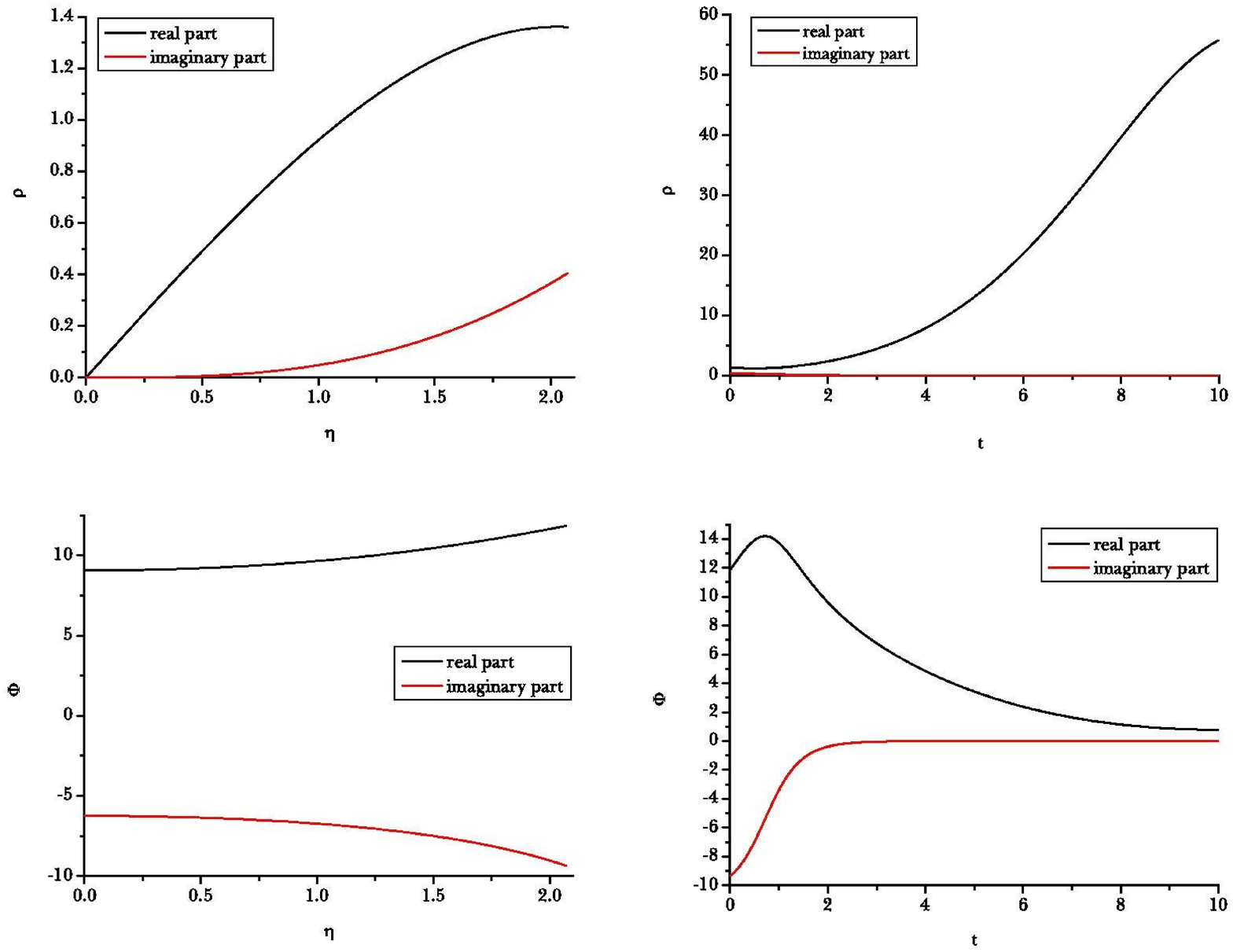}
\caption{\label{fig:quadratic_field}The classical solution $\rho^{\mathfrak{Re}}$, $\rho^{\mathfrak{Im}}$, $\Phi^{\mathfrak{Re}}$, and $\Phi^{\mathfrak{Im}}$ as functions of $\eta$ and $t$, for $\omega=-1$, $M=0.75$, and $\Phi_{0}=11$. In this case, the search algorithm determined the turning time to be $X = 2.0734$ and the initial phase angle to be $\theta = 5.68017$.}
\end{center}
\end{figure}

We have numerically determined the classical histories contributing to the no-boundary condition in the case of the quadratic potential. Our algorithm converges well and finds a region of initial values for which there are classical histories. In this region one solution per initial value  $\Phi_{0}=|\Phi(\eta=0)|$ is found. This is in contrast to the Einstein case, where there are two solutions due to the symmetry of the potential.
under $\phi\mapsto -\phi$. Figure~\ref{fig:quadratic_field} shows an example of a classical solution in Euclidean and Lorentzian time, respectively.
In Euclidean time, the real part of the scale factor $\rho$ increases as a sine function, which is typical of a de Sitter-like space in the Euclidean signature. The real part of the field $\Phi$ slowly increases, since for Euclidean signature, the potential is effectively inverted. After the turning point, the real part of the scale factor increases exponentially as is typical when there is positive vacuum energy. $\Phi$ rolls down to the equilibrium near $\Phi=1$, as expected. Other classical histories show similar characteristics, as we had already anticipated by analytical arguments.

\begin{figure}
\begin{center}
\includegraphics[scale=0.3]{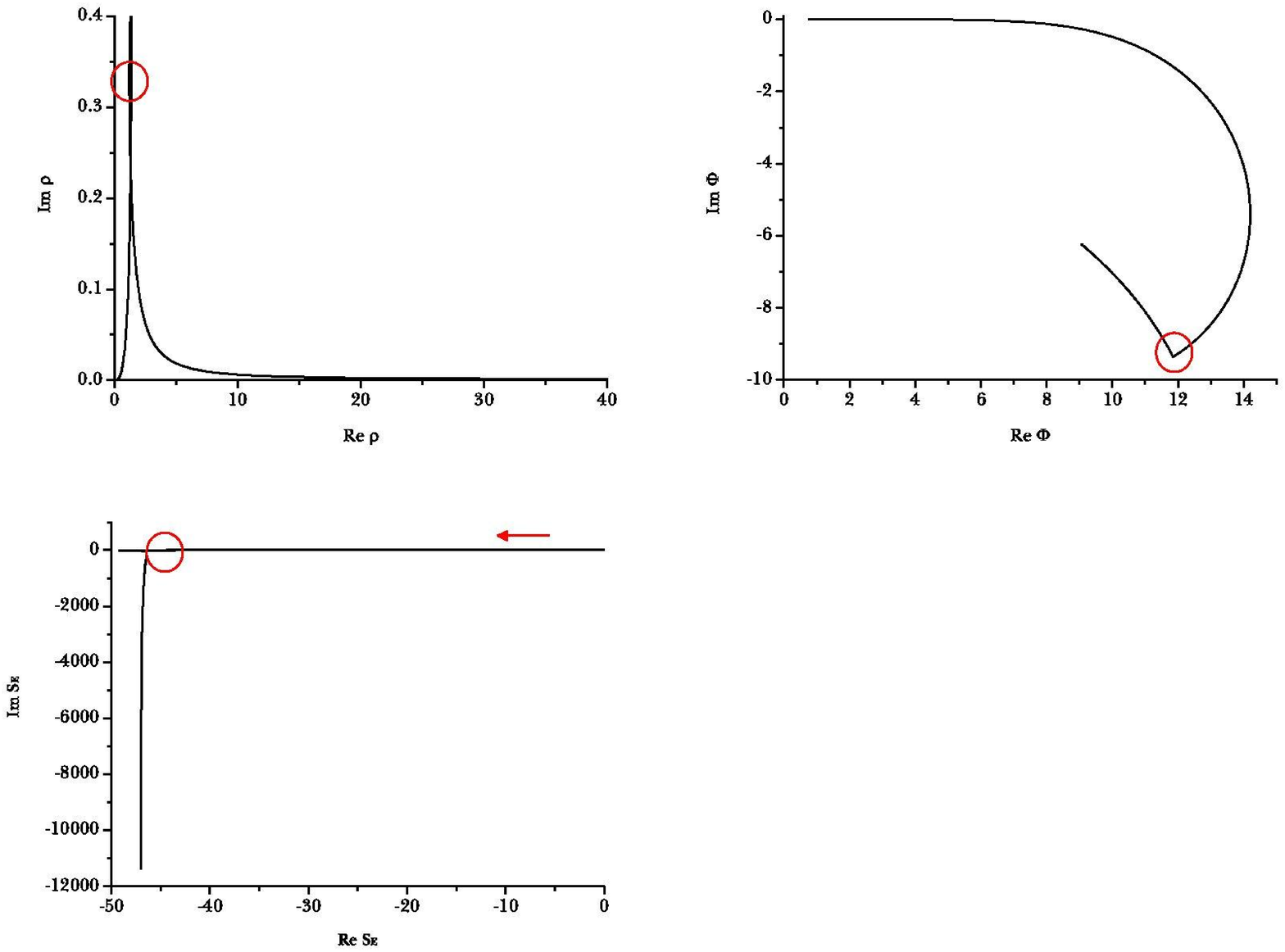}
\caption{\label{fig:quadratic_field2}Plots for $\rho^{\mathfrak{Re}}$-$\rho^{\mathfrak{Im}}$, $\Phi^{\mathfrak{Re}}$-$\Phi^{\mathfrak{Im}}$, and $S_{\mathrm{E}}^{\mathfrak{Re}}$-$S_{\mathrm{E}}^{\mathfrak{Im}}$ for $\omega=-1$, $M=0.75$, and $\Phi_{0}=11$. Red circles are the turning point.}
\end{center}
\end{figure}

We also note that the solution becomes real in the large $t$ limit, as it should: It is instructive to  compare the real part and the imaginary part (top and middle of Figure~\ref{fig:quadratic_field2}). After the turning point (marked by the red circles), the imaginary parts quickly decrease to zero and are in particular extremely small as compared to the real parts.

Also the behavior  of the action (bottom of Figure~\ref{fig:quadratic_field2}) is consistent with our expectations. During the Euclidean time, the real part of the Euclidean action increases (to the negative direction) and variation of the imaginary part is negligible. However, after the turning point, the real part of the Euclidean action is almost constant, while the imaginary part of the action increases (to the negative direction). This shows that for large $t$, the variation of the imaginary part of the action is much bigger than that of the real part, and hence the classicality condition holds for the solution. The value of the Euclidean action at late time $t$ can then be used to determine the probability of such a history: $P \sim \exp(-2S_{\mathrm{E}})$.
\begin{figure}
\begin{center}
\includegraphics[scale=1]{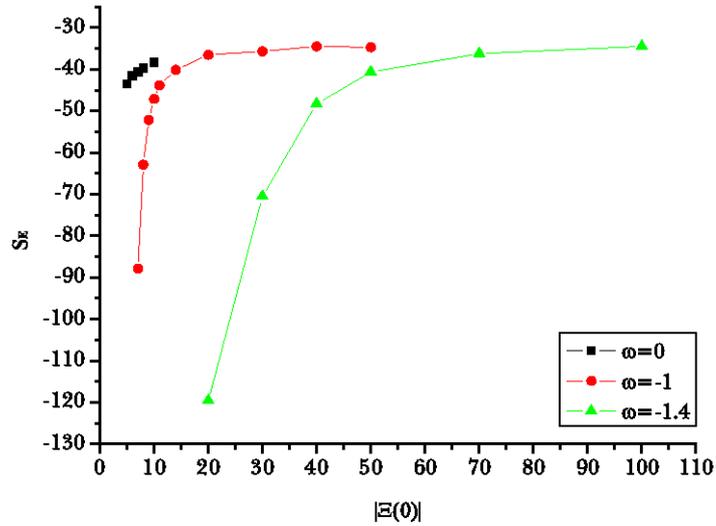}
\caption{\label{fig:quadratic}Euclidean action $S_{E}$ for quadratic potentials with $\omega=0, -1, -1.4.$}
\end{center}
\end{figure}

In Figure~\ref{fig:quadratic}, we plot the Euclidean action as a function of $|\Xi(0)|$ for several values of $\omega$. The behavior is  as anticipated in the previous section: as $\Phi_{0}$ increases, probability decreases. For a given $\omega$, there is a critical $\Phi_{0}$ so that there are no classical histories with initial value less than the critical one, which confirms our discussion using the slow roll conditions.
As we decrease $\omega$, classical histories can be observed in the small $\Phi_{0}$ region. This is related to the effective mass $m^{2}=M^{2}/(2\omega+3)$. If the effective mass increases, or $\omega \gtrsim -1.5$, then the field rolls more quickly than in the small effective mass cases, and hence it will be difficult to see a classical history near the minimum.

\subsection{\label{sec:mul}Double-well potential}

The double-well potential has two minima for $\Phi$; this is very important for the present work since it gives a simple model that allows two different universes with different gravitational couplings. After we gain intuition on the no-boundary measure in this case, we can apply it to the dilaton potential, by shifting one minimum to infinity.
\begin{figure}
\begin{center}
\includegraphics[scale=0.2]{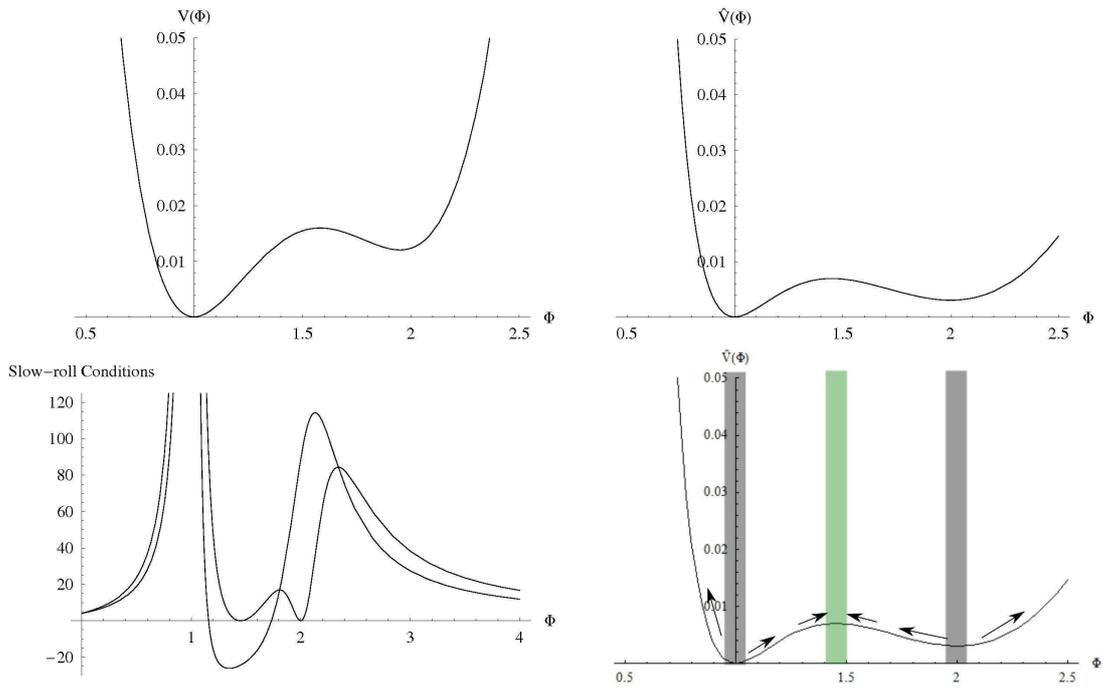}
\caption{\label{fig:potential_all}A typical double-well potential in the Jordan frame (upper left), Einstein frame (upper right), and slow-roll conditions (Equation~(\ref{eqn:SL}), the second of them positive definite) of the potential (lower left), for $A=1$, $\Phi_{\mathrm{a}}=1$, $\Phi_{\mathrm{b}}=2$, $\delta=-0.05$, and $V_{0}=0.0001$. Lower right: The regions near the local minima (black regions) do not attract trajectories during the Euclidean evolution. The local maximum (green region) is an attractor during Euclidean time.}
\end{center}
\end{figure}

\subsubsection{Analytic considerations}
Let us consider a simple case
\begin{eqnarray}
F(\Phi) &\equiv& \Phi V'(\Phi) - 2 V(\Phi) \\
&=& A \left(\Phi-\Phi_{\mathrm{a}} \right) \left(\Phi-\Phi_{\mathrm{b}} \right) \left(\Phi- \left( \frac{\Phi_{\mathrm{a}}+\Phi_{\mathrm{b}}}{2} + \delta \right) \right),
\end{eqnarray}
and $A=1$, $\Phi_{\mathrm{a}}=1$, $\Phi_{\mathrm{b}}=2$, $\delta=-0.05$, and $V_{0}=0.0001$. We plot the potential in the Jordan frame $V(\Phi)$ (upper left of Figure~\ref{fig:potential_all}), potential in the Einstein frame $\hat{V}(\Phi)$ (upper right of Figure~\ref{fig:potential_all}), and both of slow-roll conditions (lower left of Figure~\ref{fig:potential_all}). One can see that the slow-roll conditions never become sufficiently small at the same time. Perhaps, if a saddle point and an inflection point coincide in the Einstein frame, it may be possible to obtain a region where both slow-roll conditions hold. However, in this case the solution would not be stable and  hence we will not consider this possibility further.

However, even though slow-roll conditions do not hold, if initial conditions of the field are finely tuned so that if the field is slowed down and approaches a local extremum, there may be a possibility to see a classical history. One trivial example of this is the solution in Equation~(\ref{eq:exact}). Therefore, one may guess that such a real solution can happen near the solution of $\Phi V'(\Phi)- 2V(\Phi)=0$ (local minimum or local maximum of the potential in the Einstein frame).

However, if the solution is a minimum, this can not work. If the field does not start exactly at the minimum,
it will roll to the upper region of the potential during the Euclidean time evolution. Only a local maximum is an attractor during Euclidean time (lower right of Figure~\ref{fig:potential_all}), and may thus provide for a continuum of classical histories via false vacuum inflation.
Thus there is some hope to see classical solutions
\begin{enumerate}
\item precisely at a local minimum,
\item precisely at a local maximum, and
\item for initial conditions \textit{near} the local maximum.
\end{enumerate}

Now let us define subsets of all histories $\mathfrak{H}$ by imposing \textit{initial conditions}: $\mathfrak{H}_{A_{1}}$, $\mathfrak{H}_{A_{2}}$, and $\mathfrak{H}_{A_{3}}$, such that
\begin{eqnarray}
\mathfrak{H}_{A_{1}} &=& \{h \;|\; \Phi_{0} = \Phi_{m} \},\\
\mathfrak{H}_{A_{2}} &=& \{h \;|\; \Phi_{0} = \Phi_{M} \},\\
\mathfrak{H}_{A_{3}} &=& \{h \;|\; \Phi_{0} > \Phi_{M} - \Delta \Phi^{(1)} \mathrm{\;\;or\;\;} \Phi_{0} < \Phi_{M} + \Delta \Phi^{(2)} \},
\end{eqnarray}
where $\Phi_{m}$ is a local minimum, $\Phi_{M}$ is a local maximum, and $\Delta \Phi^{(1)}$, $\Delta \Phi^{(2)}>0$ are bounds to allow classical histories around the local maximum.
We know that around the local minimum or maximum, the Euclidean action will be approximately Equation~(\ref{eqn:dS}), and hence
\begin{eqnarray}
\label{eqn:PsiPsi1}\frac{P_{A_{1}}}{P_{A_{3}}} &\simeq& \exp\left(\frac{6 \pi \Phi_{m}^{2}}{V(\Phi_{m})} - \frac{6 \pi \Phi_{M}^{2}}{V(\Phi_{M})}\right) \frac{\mu (\mathfrak{H}_{A_{1}})}{\mu(\mathfrak{H}_{A_{3}})}, \\
\label{eqn:PsiPsi2}\frac{P_{A_{2}}}{P_{A_{3}}} &\simeq& \frac{\mu (\mathfrak{H}_{A_{2}})}{\mu(\mathfrak{H}_{A_{3}})},
\end{eqnarray}
where $\mu$ is given by the Lebesgue in the space of initial conditions. Therefore probabilities will be dominated by $P_{A_{3}}$, since sets $\mathfrak{H}_{1}$ and $\mathfrak{H}_{2}$ are measure zero sets.

Now let us focus on the Case $(iii)$. Then there may be two possibilities: after the history has become real-valued, it can roll down to left or right in the double-well potential. Let us impose the final conditions $B_{L}$ and $B_{R}$ and define subsets $\mathfrak{H}_{B_{L}}$ and $\mathfrak{H}_{B_{R}}$ by
\begin{eqnarray}
\mathfrak{H}_{B_{L}} &=& \{h \;|\; \Phi(\lambda=1) = \textrm{left\, side}\},\\
\mathfrak{H}_{B_{R}} &=& \{h \;|\; \Phi(\lambda=1) = \textrm{right\, side}\},
\end{eqnarray}
where $\lambda=1$ means a sufficiently large time along the Lorentzian direction.
If the potential \textit{near} the local maximum is symmetric, then for given $\Phi_{0}$, both of histories will be allowed. In other words, after the field is slowed-down at the top of the hill, if the velocity of the field is almost zero, then there will be no principle to push the field left or right. Therefore, as long as the potential is approximately symmetric near the local maximum, if a left-rolling solution is allowed, then there will be a right-rolling solution, too. In other words,
\begin{eqnarray}
\frac{P_{A_{3}\rightarrow B_{L}}}{P_{A_{3}\rightarrow B_{R}}} \simeq \frac{\mu (\mathfrak{H}_{A_{3}\rightarrow B_{L}})}{\mu(\mathfrak{H}_{A_{3}\rightarrow B_{R}})} \simeq \frac{\int_{h \in \mathfrak{H}_{A_{3}\rightarrow B_{L}}} d\Phi_{0}}{\int_{h \in \mathfrak{H}_{A_{3}\rightarrow B_{R}}} d\Phi_{0}} \simeq \mathcal{O}(1),
\end{eqnarray}
and there is no exponential contribution to determine left or right.

\begin{figure}
\begin{center}
\includegraphics[scale=0.25]{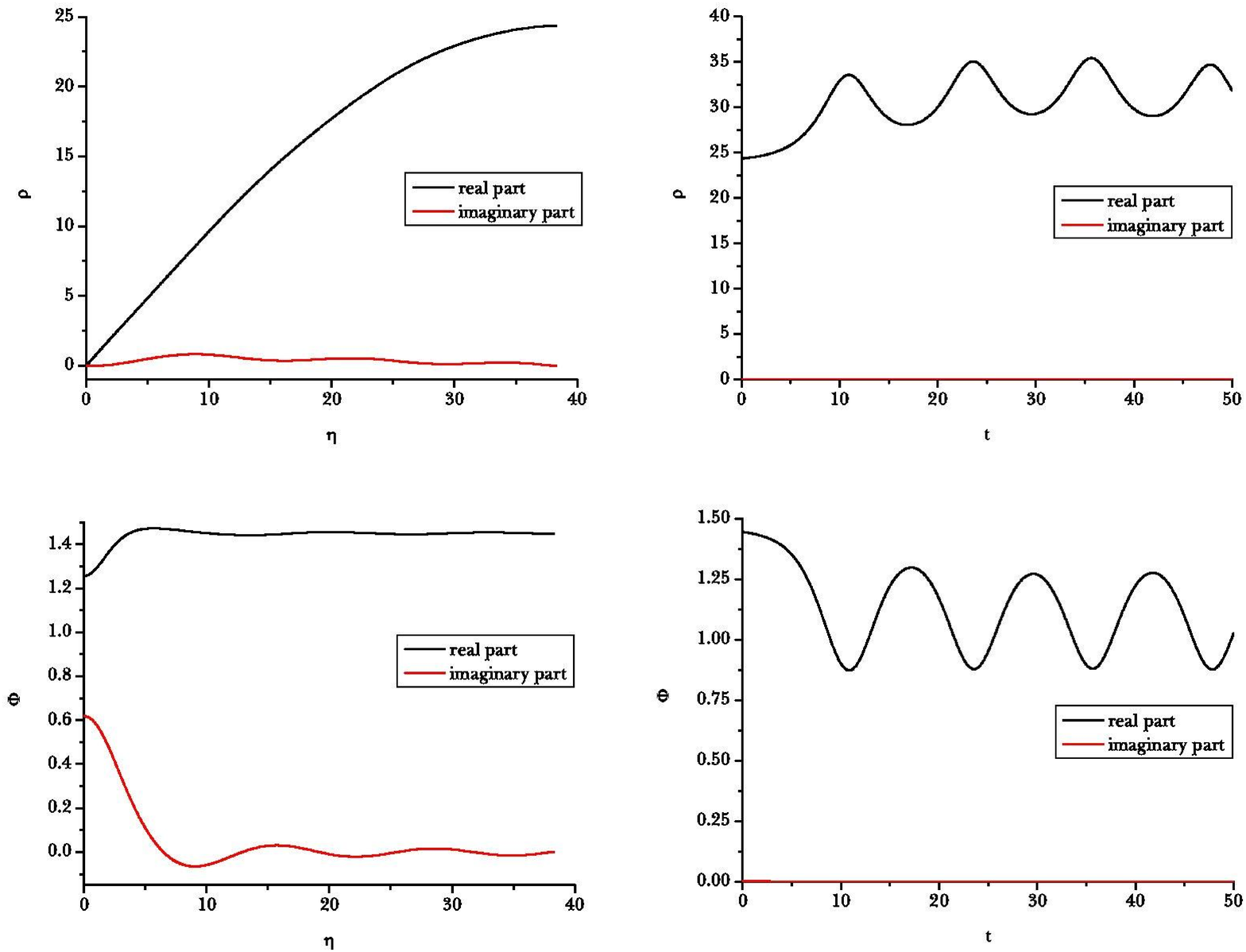}
\caption{\label{fig:left}An example of left-rolling solution $\rho^{\mathfrak{Re}}$, $\rho^{\mathfrak{Im}}$, $\Phi^{\mathfrak{Re}}$, and $\Phi^{\mathfrak{Im}}$ as functions of $\eta$ and $t$, for $A=1$, $\Phi_{\mathrm{a}}=1$, $\Phi_{\mathrm{b}}=2$, $\delta=-0.05$, $V_{0}=0.0001$, and $\Phi_{0}=1.4$.}
%\end{center}
%\end{figure}
%\begin{figure}
%\begin{center}
\includegraphics[scale=0.25]{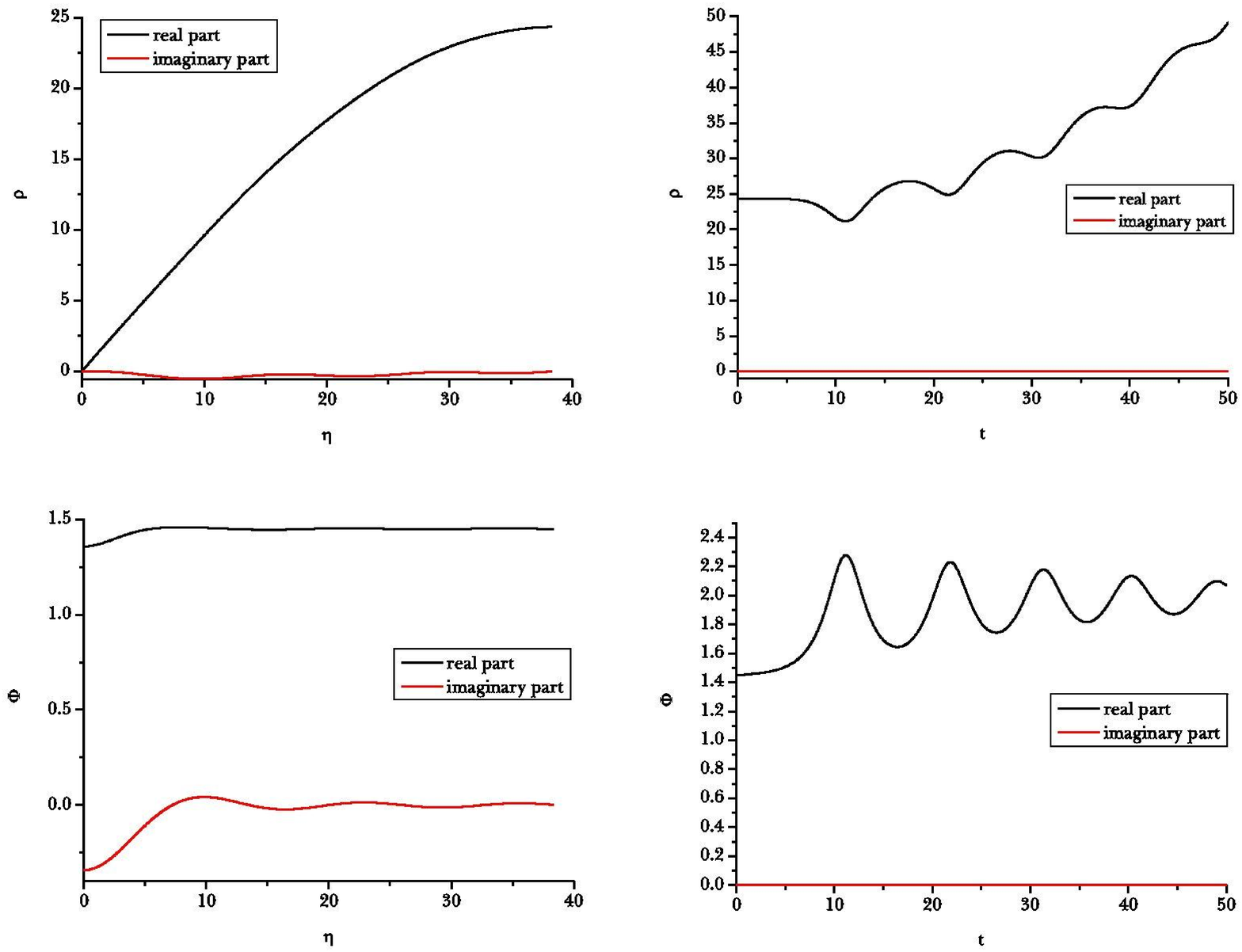}
\caption{\label{fig:right}An example of right-rolling solution $\rho^{\mathfrak{Re}}$, $\rho^{\mathfrak{Im}}$, $\Phi^{\mathfrak{Re}}$, and $\Phi^{\mathfrak{Im}}$ as functions of $\eta$ and $t$, for $A=1$, $\Phi_{\mathrm{a}}=1$, $\Phi_{\mathrm{b}}=2$, $\delta=-0.05$, $V_{0}=0.0001$, and $\Phi_{0}=1.4$.}
\end{center}
\end{figure}

\subsubsection{Numerical confirmations}

We have numerically confirmed these theoretical assertions. For convenience, we choose parameters to almost symmetric near the local maximum: $A=1$, $\Phi_{\mathrm{a}}=1$, $\Phi_{\mathrm{b}}=2$, $\delta=-0.05$, and $V_{0}=0.0001$. This potential has local minima around $\Phi=1$ and $\Phi=2$ and the local maximum is $\Phi\simeq 1.45$. Figure~\ref{fig:left} is an example of the left-rolling solutions and Figure~\ref{fig:right} is an example of the right-rolling solutions. In both cases, we fixed $\Phi_{0}=1.4$. For the right-rolling case, since the final state has sufficient vacuum energy, one can see the exponentially increasing $\rho$; therefore, future evolutions of left-rolling and right-rolling cases are quite different. This implies that for a given initial field amplitude $\Phi_{0}$, there are two physically different solutions (with different $\theta$ and $X$). Therefore, after we fix the both of initial and final conditions, $\Phi_{0}$ points out a unique history. In both cases, along the Euclidean time, the imaginary parts are definitely suppressed.

\begin{figure}
\begin{center}
\includegraphics[scale=1]{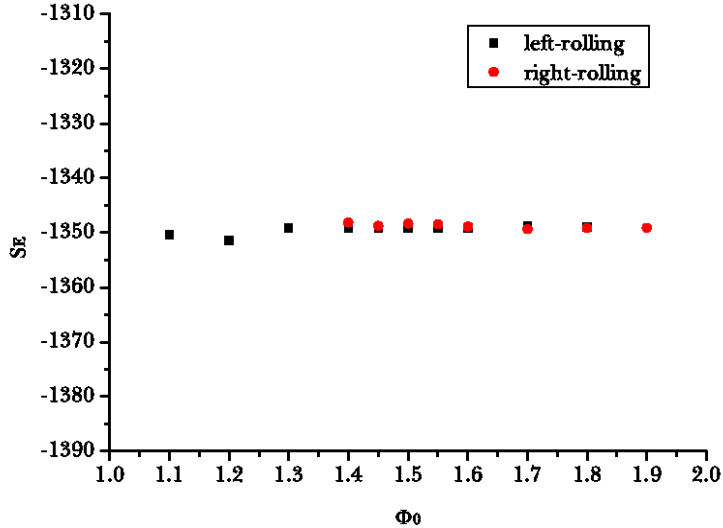}
\caption{\label{fig:doublewell}Euclidean action $S_{\mathrm{E}}$ for double-well potential: $A=1$, $\Phi_{\mathrm{a}}=1$, $\Phi_{\mathrm{b}}=2$, $\delta=-0.05$, and $V_{0}=0.0001$.}
\end{center}
\end{figure}

As we vary $\Phi_{0}$, we can classify left-rolling solutions and right-rolling solutions and estimate Euclidean actions. The allowed region for left-rolling solutions is $\Phi_{0} \lesssim 1.8$ and the allowed region for right-rolling solutions is approximately $\Phi_{0} \gtrsim 1.4$ (Figure~\ref{fig:doublewell}). Moreover, there is no meaningful difference on actions between various $\Phi_{0}$ and left or right-rolling processes. The values are approximately $-1349$, where $-3 \pi 1.45^{2}/V(1.45) \simeq -1348.9$. This confirms our assertions on the probability (up to the overall normalization factor $Z$):
\begin{eqnarray}
P_{A_{3}\rightarrow B} \simeq \exp{ \frac{6\pi \Phi_{M}^{2}}{V(\Phi_{M})} } \int_{h \in \mathcal{H}_{A_{3}\rightarrow B}} d\Phi_{0}
\end{eqnarray}
and
\begin{eqnarray}
\frac{P_{A_{3}\rightarrow B_{L}}}{P_{A_{3}\rightarrow B_{R}}} \simeq \frac{\int_{h \in \mathcal{H}_{A_{3}\rightarrow B_{L}}} d\Phi_{0}}{\int_{h \in \mathcal{H}_{A_{3}\rightarrow B_{R}}} d\Phi_{0}} \simeq \mathcal{O}(1),
\end{eqnarray}
where $\Phi_{M}$ is a local maximum in the Einstein frame.

In summary, we find in all the potentials that we have investigated numerically, that if there is a local maximum between two local minima, then the probability is determined by the local maximum.

\subsubsection{Generalization to the multiple-well potential}

\begin{figure}
\begin{center}
\includegraphics[scale=0.5]{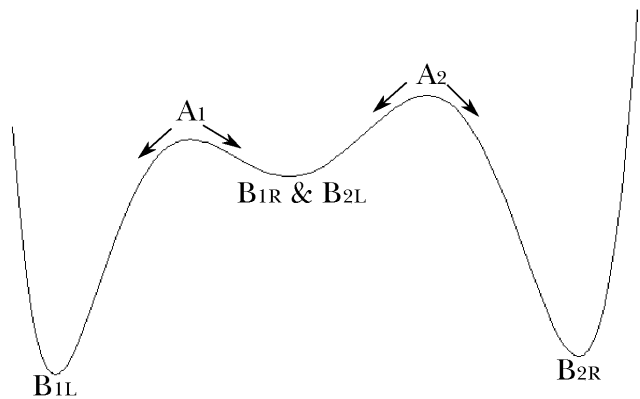}
\caption{\label{fig:multiplewell}Multiple-well potential.}
\end{center}
\end{figure}

Now let us generalize for multiple-well cases. If we see a triple-well potential, then there are two places with non--zero measure where classical solutions can be obtained: let us call $\Phi_{A_{1}}$ and $\Phi_{A_{2}}$ (Figure~\ref{fig:multiplewell}). Then, there can be three final conditions: ends at the first minimum $\mathfrak{H}_{1}$, ends at the second minimum $\mathfrak{H}_{2}$, and ends at the third minimum $\mathfrak{H}_{3}$. Histories for $\mathfrak{H}_{1}$ should begin from $A_{1}$, and each history contributes the probability approximately $\exp{(6 \pi \Phi_{A_{1}}^{2}/V(\Phi_{A_{1}}))}$. Histories for $\mathfrak{H}_{2}$ can begin from $A_{1}$ or $A_{2}$ and each history contributes the probability approximately $\exp{(6 \pi \Phi_{A_{1}}^{2}/V(\Phi_{A_{1}}))}$ or $\exp{(6 \pi \Phi_{A_{2}}^{2}/V(\Phi_{A_{2}}))}$. Finally, histories for $\mathfrak{H}_{3}$ can begin from $A_{2}$ and each histories contribute approximately $\exp{(6 \pi \Phi_{A_{2}}^{2}/V(\Phi_{A_{2}}))}$. Now we write probabilities by (we ignore the prefactor $1/Z$),
\begin{eqnarray}
P_{\mathfrak{H}_{1}} &\simeq& \exp{\frac{6 \pi \Phi_{A_{1}}^{2}}{V(\Phi_{A_{1}})}} \int_{h \in \mathfrak{H}_{A_{1}\rightarrow B_{1L}}} d\Phi_{0},\\
P_{\mathfrak{H}_{2}} &\simeq& \exp{\frac{6 \pi \Phi_{A_{1}}^{2}}{V(\Phi_{A_{1}})}} \int_{h \in \mathfrak{H}_{A_{1}\rightarrow B_{1R}}} d\Phi_{0} + \exp{\frac{6 \pi \Phi_{A_{2}}^{2}}{V(\Phi_{A_{2}})}} \int_{h \in \mathfrak{H}_{A_{2}\rightarrow B_{2L}}} d\Phi_{0},\\
P_{\mathfrak{H}_{3}} &\simeq& \exp{\frac{6 \pi \Phi_{A_{2}}^{2}}{V(\Phi_{A_{2}})}} \int_{h \in \mathfrak{H}_{A_{2}\rightarrow B_{2R}}} d\Phi_{0}.
\end{eqnarray}
This argument can be generalized for other multiple-well potentials.

\subsection{\label{sec:dil}Stabilization problem: dilaton-type potential}

\begin{figure}
\begin{center}
\includegraphics[scale=0.5]{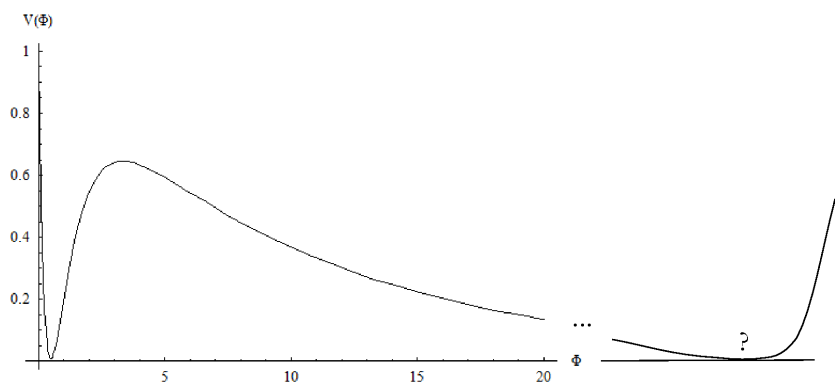}
\caption{\label{fig:dilaton}A typical dilaton-type potential and an extension for a thought experiment.}
\end{center}
\end{figure}

Let us think a typical dilaton-type potential (Figure~\ref{fig:dilaton}). In many models of dilaton potential, there is an unstable direction in the large $\Phi$ limit. Let us try to consider this as an extreme limit of a double-well potential, as in Figure~\ref{fig:dilaton}. Let us call the minima $\Phi_{m_{1}}$ and $\Phi_{m_{\infty}}$. As we have seen previously, histories in which $\Phi$ is precisely located in one of the minima will lead to classical points of the no-boundary wave function with modulus squared proportional to $\exp 6 \pi \Phi_{m}^{2}/V(\Phi_{m})$. Therefore, if we compare just these two points, the right minimum $\Phi_{m_{\infty}}$ is much more preferred. We can think that a dilaton potential is an extreme limit of $\Phi_{m_{\infty}} \rightarrow \infty$. In this point of view, Euclidean quantum cosmology seems to say that the dilaton field should be destabilized and all coupling constants of nature should be zero.


However, as we discussed in the previous section, other histories also contribute classical points, in particular  the \textit{fuzzy instantons}.
For the potential in Figure~\ref{fig:dilaton}, it is also difficult to find a region that allows slow-roll conditions to be satisfied, except inflection points. Thus there can be classical solutions around the local minima or maxima; those starting near but not precisely at the local minimum will evolve towards the maximum and obtain the main contribution to the Euclidean action there. Furthermore there will be continuously many classical histories starting around the local maximum of the potential. Therefore, as in Equations~(\ref{eqn:PsiPsi1}) and (\ref{eqn:PsiPsi2}), the local minima will be excluded.

For obvious practical reasons, we cannot directly work with the $\Phi = \infty$ solution itself. Numerically, we only see runaway solutions. However, as we numerically observed, all runaway solutions should experience false vacuum inflation around the local maximum, and hence the probability cannot be $\exp{\infty}$.

Then, are the stable histories preferred? We have to compare the left-rolling (stable) histories and the right-rolling (unstable) histories. As we discussed above, our numerical results and the understanding of the behavior of the histories contributing to the saddle point approximation that we have developed in the previous sections show that the probabilities for a classical universe with stabilized, and one with non-stabilized couplings should be of similar magnitude. At very least, it appears that the stabilized universe would not be exponentially suppresed, unless the potentially would somehow be extremely (``exponentially'') asymetric.

In conclusion, we can say on the dilaton stabilization problem using the no-boundary measure that:
\begin{enumerate}
\item We \textit{disagree} with the assertion that the probability of a universe corresponding to the runaway solution is $1$.
\item Rather, under the approximation we are working in, probabilities for stabilization and de-stabilization are of similar order.
\end{enumerate}
Therefore, the no-boundary proposal seems to be able to contribute to the solution of the dilaton stabilization problem.

\section{\label{sec:dis}Discussion}

In this paper, we investigated the no-boundary measure in scalar-tensor gravity in the context of Euclidean quantum cosmology. Especially, we were interested in trying to explain why we see a non-vanishing gravitational
coupling. This is related to the dilaton stabilization problem, to explain why the dilaton field is located in a stable vacuum of a potential.

To get a handle on this problem, we worked with a Brans-Dicke field with various potentials. As was found in previous work, we have seen that it is important to not only consider real field configurations in the saddle point approximation to the wave function, but also complex ones, the fuzzy instantons. In our case, we see that fuzzy instantons contributing to the probability for classical universes are allowed in the following two regions in a potential: With the scalar \emph{at} a local minimum or \textit{near} a local maximum. The instantons at the minimum is in fact not fuzzy, but they are of measure zero as compared to the other fuzzy instantons. Therefore, even though the former have large negative Euclidean action, their probability will be zero.

Thus, histories contributing to the probability for a classical universe
should slow down and spend time around a local maximum, experience false-vacuum inflation, and then turn to Lorentzian time. \textit{This is indeed a new classicalization mechanism} that is different from the previous work of Hartle, Hawking, and Hertog.

The next question is whether we will roll-down to left or to right. In at the level of precision and approximation that we are working at, we cannot decide. Probabilities of two possibilities have similar order, unless the potential is extraordinarily asymmetric. Therefore, the no-boundary measure can partly explain the stabilization of some coupling constants of nature. But it does not seem to assign probability zero universes in which they run away. This is perhaps a point where anthropic reasoning may be employed.

There are some issues that we have not studied, but which may nevertheless be relevant to the questions we considered:
\begin{itemize}
\item In this work, we did not include volume weighting, since we did not include an inflaton field (Of course, the dilaton field can role as an inflaton field, but it may not necessarily be true). If we include the \textit{inflaton field}, the classicalization process could be changed, since there is another field that induces inflation. The inclusion of volume weighting and/or an additional scalar may change the results.

\item If there is a correction term which affects histories near a local minimum so that it \textit{breaks symmetry} around the local minimum, then the region near the local minimum could admit a continuous spectrum of histories. Then, if vacuum energy of such a local minimum is sufficiently smaller than other positions, it may fully explain the stabilization of the dilaton field.

\item From our numerical results it seems that our intuition, coming from the dynamics of real fields, serves us well for the qualitative understanding of  the dynamics of the fuzzy instantons. It should however be kept in mind that these are complex field configurations. In particular, they are sensitive to the analytic continuation of the potential. It is thus not inconceivable that different potentials, with a similar shape on the real sub-sector of the theory, will lead to different results.
\end{itemize}

In addition, although our study is on scalar-tensor gravity, our conclusion is qualitatively relevant also for the no-boundary measure of Einstein gravity minimally coupled to a scalar field. Due to their choice of the potential, Hartle, Hawking and Hertog \cite{Hartle:2007gi,Hartle:2008ng} only considered slow-roll inflation, concluded that the bottom-up probabilities do not favor larger amounts of inflation, and then appealed to volume weighting to deal with this problem \cite{Hawking:2002af}. Our work shows, however, that there is a second mechanism to generate classical histories, provided the potential has a more complex shape. They can be formed around the local maximum via false vacuum inflation. In this case, the bottom-up probability does not disfavor the top of the hill of the potential, and it may be possible to explain large amounts of inflation from bottom up probabilities in this way.

\section*{Acknowledgment}
The authors would like to thank Bum-Hoon Lee and Ewan Stewart for discussions and encouragement. We also thank to Min-jae Kim and Sun-young Lee for discussions on optimization algorithms, and an editor and a referee at CQG for many comments that substantially improved the manuscript. DY and DH were supported by Korea Research Foundation grants (KRF-313-2007-C00164, KRF-341-2007-C00010) funded by the Korean government (MOEHRD) and BK21. DY was supported by the National Research Foundation of Korea(NRF) grant funded by the Korea government(MEST) through the Center for Quantum Spacetime(CQUeST) of Sogang University with grant number 2005-0049409. HS would like to thank Bum-Hoon Lee for hospitality at the Center for Quantum Spacetime(CQUeST) of Sogang University where part of this work was completed. His work was partially supported by the Spanish MICINN project No. FIS2008-06078-C03-03.

\end{document}